\documentstyle[12pt]{article}
\begin{document}
\title{New Mechanism of Collapse and Revival in Wave Packet Dynamics
Due to Spin--Orbit Interaction
\thanks{Presented at the XXXI Zakopane School of Physics,
Zakopane, Poland, September 3 -- 11, 1996.} }
\author{\underline{P.~Rozmej}$^{\rm a}$,
W.~Berej$^{\rm a}$ and R.~Arvieu$^{\rm b}$ \\ \small
$^{\rm a}$Theoretical Physics Department, University MCS, 20-031 Lublin, Poland\\
\small rozmej@tytan.umcs.lublin.pl\\ \small
$^{\rm b}$Institut des Sciences Nucl\'eaires, F 38026 Grenoble-Cedex, France\\
\small arvieu@frcpn11.in2p3.fr}

\date{November 14, 1996}
\maketitle

\begin{abstract}
This article discusses the properties of time evolution of wave packets
 in a few systems.
Dynamics of wave packet motion for Rydberg atoms
with the hierarchy of collapses and revivals is briefly reviewed.
The main part of the paper focuses on the new mechanism of quantum
recurrences in wave packet dynamics. This mechanism can occur in any
physical system with strong enough spin--orbit interaction.
We discuss here the {\em spin--orbit pendulum} effect that consists
in different motions of subpackets with different spin fields and
results in oscillations of a fraction of average angular momentum
between spin and ordinary subspaces. The evolution of localized wave
packets into toroidal objects and backwards (for other class of initial
conditions) is also subject to discussion.

\vspace*{2mm}\noindent
PACS number(s): 03.65.Sq, 03.65.Ge, 32.90+a
\end{abstract}

\section{Introduction}
Recent advances in experimental techniques have resulted in rapidly
growing attention to the dynamics of particular nonstationary
states, wave packets. With tunable, short and intensive laser pulses
a creation and analysis of such states in atomic and molecular systems
became available. An extensive research by means of both theoretical
and experimental methods has brought an understanding of the intriguing
phenomena of a hierarchy of collapses and revivals for particularly
prepared wave packets in Rydberg atoms
$[$1-7$]$ 
and in Jaynes--Cummings model (JCM) of quantum optics $[$8-12$]$.

In hydrogen and hydrogenoid atoms electrons can be excited to a coherent
mixture of many Rydberg states that move almost classically for many
Kepler periods $[$1-5,7$]$.  
However due to a nonequidistant spectrum of energy levels wave
packets undergo a sequence of spreadings, collapses and revivals during long
term evolution. There exist several time scales for the revivals
(reccurences). The shortest one is associated with the period of a classical
motion, i.e. Kepler period, the other ones called $t_{rev}$ and $t_{sr}$
(revival and superrevival times respectively) arise from subtle quantum
interference effects. The examples of such effects are described in more
detail in Section 2.

In atomic systems the appropriate wave packet can be excited by a suitably
chosen laser pulse $[$2,6$]$. 
The time evolution can be then probed
by the following pulse after precisely measured time delay.
Using additionally a circularily polarized electric field allows for
a creation of electronics wave packets which are stationary  $[$13,14$]$.
In this case the wave packet motion is stabilized by a nonlinear coupling
in the analogous way as orbits of Trojan asteroids are
stabilized in the celestial mechanics $[$15$]$. 

Another interesting phenomenon already observed in quantum optics is an
atomic analogue to a double--slit experiment $[$16$]$. 
Again in hydrogen or hydrogenoid atoms one can excite, by a pair of laser
pulses with an appropriate time delay, two wave packets localized on the
opposite sides of the orbit.
Such a state is often called a 'Schr\"odinger cat state' as it is a
mesoscopic size realization of the famous Schr\"odinger cat problem
$[$17$]$. 

In the following sections 3 and 4 we discuss the new dynamical structures
in wave packet motion. These new effects arise when the strong enough
spin--orbit interaction is present in a physical system.
In order to exhibit these effects in the most transparent way we have chosen
the simple central potential in spherical harmonic oscillator form.
This choice allows for the factorization of the evolution operator into
two parts and for considering the orbital motion and spin--orbit motion
separately, as well as for finding analytical solutions for both motions.
In section 3 we present the spin--orbit pendulum effect
$[$18-20$]$ 
which is most pronounced for circular orbits. Section 4 contains the
discussion of linear trajectories for which spin--orbit interaction spreads
(periodically) the initially spherical, well localized wave packet into
a toroidal object, with a classical trajectory being the symmetry axis
of the torus. Such behaviour of the wave packet has been presented here
for the first time.

All these experimental and theoretical achievements enrich our understanding
of the basic features of quantum mechanics.

\section{Properties of Rydberg wave packets}
The discussion presented in this section is in line of that of $[$3,5,7,22$]$.

Consider a hydrogen atom prepared in such a way that the wave function is
a superposition of bound-state eigenfunctions
\begin{equation}
\label{psirt}
\Psi(\vec{r},t) = \sum_n c_n u_n(\vec{r}) \exp{(-iE_n t/\hbar)} \;,
\end{equation}
where coefficients $c_n$ are not negligible only for rather narrow range
of $n$ around the mean value $N$.
Obviously $\int u_n^* u_s d\vec{r}=\delta_{ns}$ and $\sum_n |c_n|^2=1$.
This sum converges, therefore a finite number of $c_n$ makes it arbitrarily
close to 1 and is sufficient to represent $\Psi$ with arbitrary accuracy.
Then any $\Psi$ is arbitrarily close to a periodic function of time $[$23$]$.
Let $K$ be the common multiple of all $n$ for which $c_n$ are not
neglected. Then $\Psi$ has a period $T_{cl}K^2$.
The recurrence with this period is exact, but if many levels contribute
to the sum (1) 
the required time is enormously long.
However, nearly exact recurrences occur considerably earlier.
The assumption that weights $c_n$ are strongly centered around a mean
value $N$ permits an expansion of the energy in a Taylor series
\begin{equation}
\label{eser}
E_n \simeq E_{N} + E_{N}'\,(n-N) +
E_{N}''\,(n-N)^2 +
E_{N}'''\,(n-N)^3 + \dots \;,
\end{equation}
where each prime on $E_{N}$ denotes a derivative.
The derivative terms in (2) 
define distinct time scales that
depend on ${N}$: $T_{cl}= 2\pi N^3 ,\:
t_{rev}=(N/3+1/2)\,T_{cl} ,\: t_{sr}=(N^2/4+1/2)\,T_{cl}$,
called classical period, revival time and superrevival time, respectively.
Keeping terms through third order and disregarding an overall
time--dependent phase, one can rewrite (1) 
as
\begin{equation}
\label{psirt1}
\Psi(\vec{r},t) = \sum_n c_n u_n(\vec{r}) \exp{ \left[
-2\pi i \left( \frac{(n-N)\,t}{T_{cl}} +
 \frac{(n-N)^2\,t}{t_{rev}} +
 \frac{(n-N)^3\,t}{t_{sr}} \right) \right]}\;.
\end{equation}

The convenient measure of the degree of recurrences is the recurrence
probability (also referred to as autocorrelation function),
where $w_n=|c_n|^2$ and time units are atomic units,
\begin{equation} \label{aucf}
 P(t) = |\langle \Psi(0) | \Psi(t)\rangle |^2= \sum_n w_n \, e^{it/n^2} \; .
\end{equation}
In order to exhibit some properties of short term and long term evolution
let us construct in a standard way the so--called circular wave packet
$[$1-4$]$. 
Then we can choose
$c_n = (2\pi\sigma)^{-1/4}\,\exp[-((n-N)/2\sigma)^2]$
and $u_n(\vec{r}) = \Phi_{n,n-1,n-1}(\vec{r})$ (aligned standard hydrogenic
eigenfunction with $l=m=n-1$).
The features of the evolution of such a circular wave packet are
illustrated below with	some numerical examples.
We show the results obtained for the wave packet with
$N=60$, $\sigma=1.5$ and the summation over $n$
taken from $n_1=50$ to $n_2=70$.
For this case $t_{rev}=20.5\,T_{cl}$ , $t_{sr}=900.5\,T_{cl}$ and
the mean principal quantum number $N$  is large enough
for the fractional revivals to appear.

The short term evolution (time range 0--50$\,T_{cl}$) is displayed
in Fig.~1, where the recurrence probability (4) 
is plotted as function of time. Initial peak intensity decreases very fast,
and it is seen that after 4 revolutions wave packet is well spread
over the whole orbit. Around $t\simeq 5\,T_{cl}$ (1/4 of $t_{rev}$)
peaks appear 4 times more frequently indicating formation of
a fractional revival of 1/4 order (see also  Fig.~3b).
Around $t\simeq 7\,T_{cl}$ (1/3 of $t_{rev}$) and
$t\simeq 10\,T_{cl}$ (1/2 of $t_{rev}$) the frequency of peaks of the
autocorrelation function indicate formations of the fractional revivals
of orders 1/3 and 1/2, respectively (shapes of the wave packets
are shown in Fig.~3c and 3d).
The so--called full revivals are built at times close to
$t\simeq 20\,T_{cl}$ and  $t\simeq 40\,T_{cl}$ and a corresponding shape
of the wave packet is exhibited in Fig.~3e.

Fig.~2 displays the same autocorrelation function (4) but in a much longer
scale. The formation of recurrences corresponding to $t_{sr}$ at
$t\simeq 300$ i.e. $t=t_{sr}/3$  and to some extent to
$t \simeq 150$ i.e.  $t=t_{sr}/6$
respectively is clearly seen.  The corresponding shape of the
wave packet at first superrevival occurence is present in Fig.~3f.

\section{Spin--orbit pendulum}
In this section we discuss the properties of the time evolution of wave
packets representing a fermion moving in a spherical harmonic oscillator
potential with a strong spin--orbit coupling. The hamiltonian of the
system is then a simplified form of the single--particle Nilsson hamiltonian,
extensively used in nuclear physics $[$24$]$ 
\begin{equation} \label{ham} 
H = H_0 + V_{ls} = H_0 + \kappa\,(\vec{l}\cdot\vec{\sigma}) \;.
\end{equation}
The extensive investigation of the problem is contained in refs.
$[$18-21$]$, 
here we review briefly the main results.
In order to make the discussion simpler let us first focus attention
on a particular case of circular orbits. Explicitly, we choose
(without any loss of generality) the $Oxy$ plane as the orbit plane
and $Ox$ as the initial spin direction (initial spin in orbit's plane).
Then the initial states take the following explicit form
\begin{equation} \label{Nxwp} 
|\Psi(t=0)\rangle = |N,\vec{x}\rangle = |N\rangle\,\frac{1}{\sqrt{2}}\,
(|+\rangle + |-\rangle ) \; ,
\end{equation}
where the eigenstate of $s_x$ is expressed explicitly by the eigenstates
of $s_z$ operator ($|+\rangle$ and $|-\rangle$) and $|N\rangle$ is the
coherent state of spherical harmonic oscillator corresponding to a
circular orbit. In configuration space it has the form
\begin{equation} \label{rnc} 
\langle \vec{r}\/|N\rangle = \pi^{-\frac{3}{4}} \,
{\rm{e}}^{-\frac{1}{2}[(x-x_0)^2+y^2+z^2]}\, {\rm{e}}^{{\rm{i}}p_0y}\;.
\end{equation}
This packet has its maximum at ($x_0$,0,0) and moves along the circular
orbit if $p_0$=$x_0$=$\sqrt{N}$, $N$ --
the average value of $n$ quantum number
(in this case average values of $l_x,l_y,l_z$ are $0,0,x_0^2$=$N$,
respectively).
Note that the length of the angular momentum is also $N$
in these units.
The state $|N\rangle$ can be decomposed in the HO basis as
\begin{equation} \label{Nwp} 
|N\rangle = \sum_l \lambda_l \,|n=l,l,m=l\rangle
 = \sum_l \lambda_l \,|l\,l\,l\rangle \;,
\end{equation}
i.e.\ it is a linear combination of states with $m$=$l$ and $n$=$l$.
This particular state which combines all the partial waves and
consequently evolves during the time evolution with the spin--orbit
coupling by using all the frequencies of the spin--orbit
partners, can be taken as `pseudo' one-dimensional harmonic oscillator
coherent state if one defines the weights $\lambda_l$ in terms of
$l$ and a (continuous) real variable $N$ as a Poisson distribution
\begin{equation} \label{pois} 
|\lambda_l|^2 = {\rm{e}}^{-N}\, \frac{N^l}{l!} \; .
\end{equation}

As operators $H_0$ and $V_{ls}$ commute, the evolution operator connected with
the Hamiltonian (5) 
can be factorized as
\begin{equation}   \label{evo} 
U(t) = U_0(t) \, U_{ls}(t) = {\rm e}^{-{\rm i}tH_0}\,
{\rm e}^{{\rm -i}t(\vec{l}\cdot\vec{\sigma})} \;,
\end{equation}
where the appropriate time units are chosen to absorb $\kappa$ and $\hbar$.
Using the equality
$(\vec{l}\cdot\vec{\sigma})^2 = l^2 - (\vec{l}\cdot\vec{\sigma}) $
in the expansion of $U_{ls}(t)$ one can rearrange it as
\begin{equation}   \label{evos} 
U_{ls}(t) = {\rm e}^{-{\rm i}t(\vec{l}\cdot\vec{\sigma})} =
f(t) + g(t)(\vec{l}\cdot\vec{\sigma})  \; ,
\end{equation}
where, with the initial condition $U_{ls}(t=0)=1$,
\begin{equation} \label{fta} 
f(t) = {\rm{e}}^{{\rm{i}}\frac{t}{2}}\left[\cos\Omega\frac{t}{2}
- \frac{{\rm{i}}}{\Omega} \sin \Omega\frac{t}{2} \right] \;,
\end{equation} 
\begin{equation} \label{ftb} 
g(t) = {\rm{e}}^{{\rm{i}}\frac{t}{2}} \left[ {}
   - \frac{2{\rm{i}}}{\Omega} \sin \Omega\frac{t}{2} \right] \;.
\end{equation}
In the last equations $ \Omega$=$\sqrt{1+4l^2}$ and states $|lm\rangle$
are its eigenstates  with the eigenvalues $(2l+1)$.

The formulae (10--13) applied to (6) give the analytical form of the
wave packet as a function of time.
In the limit of high $N$ (which in practice is achieved already for
$N \simeq 10$)   we can provide a simple explanation of
the dynamics. In this limit we obtain
 $|\Psi(t)\rangle =|\Psi_+\rangle |+\rangle + |\Psi_-\rangle|-\rangle$
where the components of the spinor are given as
\begin{equation} \label{psit+}
|\Psi_+\rangle= \frac{1}{\sqrt{2}}\, {\rm e}^{-{\rm i}tH_0}\,\sum_l  \,
\lambda_l \, {\rm e}^{-{\rm i}lt} \, |l\,l\,l\rangle \;,
\end{equation}
and
\begin{equation} \label{psit-}
|\Psi_-\rangle= \frac{1}{\sqrt{2}}\, {\rm e}^{-{\rm i}tH_0}\,\sum_l  \,
\lambda_l \, {\rm e}^{{\rm i}(l+1)t} \, |l\,l\,l\rangle \;.
\end{equation}
Each component is again a coherent state that moves on
a circular orbit of radius $x_0$, with angular velocities
$\omega_0-\omega_{ls}$ and $\omega_0+\omega_{ls}$, respectively.
There is in $|\Psi_-\rangle$ an overall phase ${\rm e}^{{\rm i}t}$ which
produces at time $t=\frac{1}{2}T_{ls}=\pi$ the simple result
 $|\Psi_-\rangle= - |\Psi_+\rangle$, which means that a pure state with
spin along $-Ox$ is then observed.

These features are shown in details in Fig.~4 where the spatial motion
of the wave packet (6-8) is plotted (more precisely only evolution
under $U_{ls}$ operator (11)). The figure displays the total probability
density $|\Psi(t)|^2$ for the case $N=18$ at times
$t_i= i/8\,T_{ls}$ , $i= 0-4$. As motion is periodic the motion in the
second half of the $T_{ls}$ period is symmetric to that of the first half.
 Fig.~5 illustrates the motion in the spin
subspace. It displays time dependence of the
$\langle s_x \rangle$ and Tr$(\rho^2)$, where $\rho$ is the density matrix
reduced to the spin subspace. The oscillations of
$\langle \vec{s}\, \rangle$ are accompanied by the corresponding
oscillations of $\langle \vec{l}\, \rangle$ as the total angular
momentum is conserved in the system. This effect of a periodic
exchange of the portion of angular momentum between the spin and
the ordinary subspaces has been called {\em spin--orbit pendulum}
$[$18,19$]$. 
For analytical formulas and an extensive discussion see
$[$18,19$]$. 

\section{Time dependent vortex rings}
In this section we want to discuss the same physical system with
the hamiltonian (5) but quite different initial conditions corresponding
to linear classical motion. More precisely we consider initial states
in the form (6) but with the shape in configuration space at time $t=0$
given by
\begin{equation} \label{rnc1} 
\langle \vec{r}\/|N\rangle = \pi^{-\frac{3}{4}} \,
{\rm{e}}^{-\frac{1}{2}[(x-x_0)^2+y^2+z^2]} \;.
\end{equation}
This is the gaussian wave packet at rest centered at ($x_0$,0,0), $x_0=
\sqrt{2N}$, which in HO potential moves along $Ox$ axis
with average value of the angular momentum equal zero
(note that according to (6) spin direction is the same).
The decomposition into the HO basis is more general
\begin{equation} \label{Nwp1} 
|N\rangle = \sum_{nlm} \lambda_{nlm} \,|n l m\rangle \;.
\end{equation}
The coefficients $\lambda_{nlm}$ are known analytically for
arbitrary initial conditions (for details see eq.~(55) of ref.
$[$21$]$ 
and related discussion). Again with (10-13) and (17) 
one can derive the analytical formulas for the time evolution.
The details of this analytical formulation in terms of time dependent
partial waves will be published elsewhere $[$25$]$. 

The examples of the motion of the wave packet with the lowest value
of anglar momentum ($L=0$) are shown in Figs. 6 and 7.
Such a wave packet in absence of the spin--orbit coupling ($\kappa=0$)
moves along the piece of the straight line on $Oz$ axis.
Both figures display only this part of the evolution which is
governed by the $U_{ls}$ operator, i.e. the harmonic oscillator motion
is frozen. Both figures clearly show that the wave packet,
initially well localized spreads into a torus (more precisely
into two tori, the second with much lower density) with the classical
trajectory being the symmetry axis. This spreading is reversible
in our system due to its integrability and periodicity.
Such 'vortex rings' topology of wave packets is presented here
for the first time.
The full motion is a superposition of spin--orbit motion
and harmonic oscillator motion (10).

\section{Conclusions}

We have shown here that the presence of a spin-orbit interaction
in a single--particle hamiltonian generates a new mechanism of collapses
and revivals in wave packet dynamics. Assuming that initial state of a
fermion is prepared in a form of a gaussian wave packet (a coherent state
of harmonic oscillator) with well determined spin projection we are
able to calculate its evolution in analytical terms.
There exist some initial conditions for which formulas become compact and
transparent, and can serve as a guidance for the most general cases.
There are two extremes: the case $N=L$ which
corresponds to a circular classical orbit and the case
$L=0$ corresponding to a linear orbit.
In our simple model the constant form factor for spin--orbit coupling
allows for a factorization of the evolution operator into two commuting
parts. This fact makes possible considering both parts of the full
motion separately and understand it better.
The evolution operator $U_{ls}$, connected with the spin--orbit interaction
results in splitting a wave packet into up and down components that
evolve in a different manner. The so--called wave packet collapse reflects the
vanishing overlap between subpackets. At $(k+1/2)T_{ls}$  the subpackets
overlap again (exactly in the limit of high $N$), with
the opposite phases and the wave packet revives with spin reversed.
At times $kT_{ls}$ the subpackets overlap exactly with the same phases
and the initial state is exactly rebuilt. The sequence of collapses
and revivals is accompanied by oscillations of average values of spin
and orbital angular momentum, the so--called {\em spin--orbit pendulum} effect.
This behaviour is common for any initial condition providing that spin
is initially in orbit's plane.

In the hydrogenoid atoms it is necessary to combine the properties 
of Rydberg wave packets described in section 2 with the spin-orbit
evolution that we have derived. In this way we obtain a complicated 
interference patern in which the time scale due to the spin-orbit
interaction must be included. No simple description has emerged yet.

\end{document}